\documentclass[conference]{IEEEtran}
\usepackage{cite}
\usepackage{amsmath,amssymb,amsfonts}
\usepackage{algorithmic}
\usepackage{graphicx}
\usepackage{textcomp}
\usepackage{xcolor}

\usepackage{subcaption}
\usepackage{times}
\usepackage{epsfig}
\usepackage{amssymb}
\usepackage{multirow}
\usepackage{listings}
\usepackage{arydshln}
\usepackage{comment}
\usepackage{tikz}
\usepackage{pgfplots}
\usepackage{caption}
\graphicspath{{images/}}
\usepackage{rotating}
\usepackage{capt-of}

\begin{document}

\title{Detection of Illegal Kiln Activity During SMOG Period\\
{\footnotesize \thanks{978-1-6654-6472-7/23/\$31.00 \copyright 2023 IEEE}
}}

\author{Usman Nazir, Muhammad Awais Ather and Murtaza Taj \\
Department of Computer Science, Syed Babar Ali School of Science and Engineering\\Lahore University of Management Sciences (LUMS), Lahore, Pakistan \\
\{17030059, 18030044, murtaza.taj\}@lums.edu.pk
}

\maketitle

\begin{abstract}
Brick Kilns are known to be a leading cause of air pollution. The toxins and gaseous emissions from brick kilns leads to SMOG. It is a mixture of invisible toxic gases such as carbon monoxide ($CO$), ozone ($O_3$), sulphur dioxide ($SO_2$) and particulate matter like soot and carcinogens. We propose how low spatial resolution remote sensing data can be used to identify illegal industrial activity i.e. kiln operation particularly during winter SMOG period through heat signature values and the concentration of gases like $CO$, $NO_2$, $SO_2$ and $O_3$ in the atmosphere.

\end{abstract}
\begin{IEEEkeywords}
Google Earth Engine, Spectral properties, Brick Kiln, Sustainable Development Goals
\end{IEEEkeywords}

\section{Introduction}
Brick Kilns are the primary cause of ambient air pollution in underdeveloped countries ~\cite{haque2022impact, hussain2022brick}. Most of the brick kilns in Asia burn coal as a fuel source, resulting in SMOG and poor air quality \cite{subhanullah2022assessment}. Coal combustion, industrial and vehicular emissions, fires and reactions of these photochemical emissions are the cause of man-made SMOG \cite{SmogWiki7online}. It is mainly present in lower part of atmosphere. SMOG is a mixture of invisible toxic gases such as carbon monoxide ($CO$), ozone ($O_3$), sulphur dioxide ($SO_2$) and particulate matter like soot and carcinogens \cite{arun2022planning}. 

\begin{figure}
    \centering
    \includegraphics[width=0.5\textwidth]{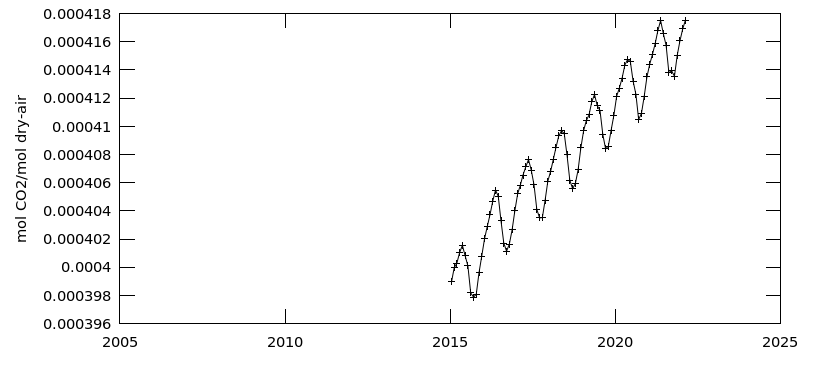}
    \caption{Greenhouse gas $CO_2$ is increasing at an exponential rate in Pakistan [Image courtesy: GIOVANNI Application].}
    \label{fig:giovanni}
\end{figure}

The emission from brick kilns mainly consists of dust particles, fine particles of coal, organic matters and low amount of gases such as $SO_2$, $NO_2$, $H_2S$ and $CO$ \cite{kiln_causes}. The emission factors per 1,000 bricks were averaged up to 6.35–12.3 kg of $CO$, 0.52–5.9 kg of $SO_2$ and 0.64–1.4 kg of particulate matter (PM) \cite{integrated}. Few of these gases are emitted directly and some are formed from the emission of air pollutants like $O_3$ from $NO_2$. That is why reduction in SMOG often involves mitigation of $NO_2$ emissions \cite{koukouli2022air}. Nonetheless, because of human activity such as burning of fossil fuels; greenhouse gases are increasing at exponential rate in atmosphere which causes the average temperature to rise and eventually leading to disastrous effects on earth's climate [see Fig.~\ref{fig:giovanni}].
\begin{figure}
    \centering
    \includegraphics[width=0.5\textwidth]{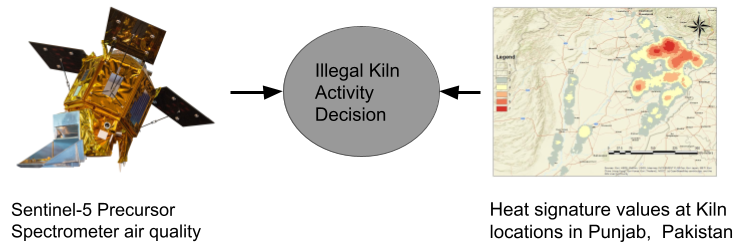}
    \caption{Proposed architecture to identify illegal kiln operation using gaseous emissions data from Sentinel-5P satellite and heat signature values.}
    \label{fig:fusion}
\end{figure}
Traditionally, production of bricks is a procedure in which hand-made bricks are baked in Bull's Trench Kilns which is about two centuries old procedure. Different other kinds of brick making plants are also introduced such as Hoffman kiln, tunnel kiln, modified FCBTK and the VSBKs. But in Pakistan, Mostly Bull's Trench Kilns with fixed chimney (FCBTK) are used. International Centre for Integrated Mountain Development (ICIMOD) introduced a new technology while working in Nepal which is based on significant improvement on Bull's Trench Kilns by modifying the flow of hot air used to progressively bake brick. This technology is known as Induced Draught Zig Zag Brick Kiln. The environment protection department (EPD) of Punjab, Pakistan in collaboration with All Pakistan Brick Kiln Owners Association is introducing this Zig Zag technology which is environment friendly and cost effective for the production of bricks. Zig Zag Kilns are said to improve fuel efficiency by 40\% and reduce 70\% gaseous emissions as compared to the conventional Bull’s Trench Kilns. 

In this study, we identify illegal kiln activities in different tehsils of Pakistan using gaseous emissions data from Sentinel-5P satellite and heat signature values [see Fig.~\ref{fig:fusion}].
The effects of air contamination have become one of the most significant test for public specialists. The measurement of emissions and their spatial appropriation are fundamental for any air quality program. We performed regional examination by utilizing a Geographic Information System (GIS). 
This advanced step  gives land use maps which can be inferred with a few characterization methods on satellite images or computerized aerial imagery. This combination of land use maps along with concentration maps of contaminants, permits acknowledgment of regions presented to high contamination levels and the relative subjection during time. The progressive step of this regional analysis permits recognizable proof of destinations ideal for the establishment of monitoring stations adhering to the standards given by the mandates in force. This methodology can further support planning of monitoring network systems concentrated on air quality. 




\begin{figure}[h]
    \centering
    \includegraphics[width=0.45\textwidth]{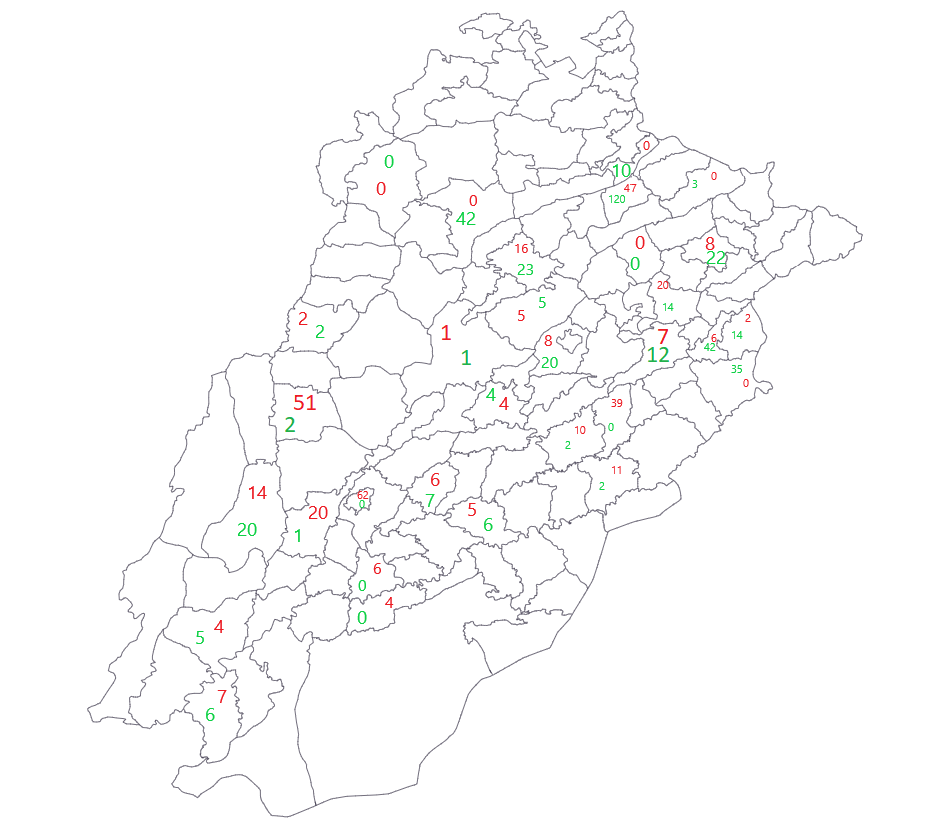}
    \caption{Ground Truth: illegal kiln activity cases by Environmental Protection Authority (EPA) in different tehsils of Punjab, Pakistan (Red number shows the No. of FIR lodged against brick kilns and green number shows the number of brick kilns sealed by EPA).}
    \label{fig:GroundTruth}
\end{figure}

 \begin{figure}[h]
 \centering
	\begin{tabular}{ccc}
			\includegraphics[width=.3\columnwidth]{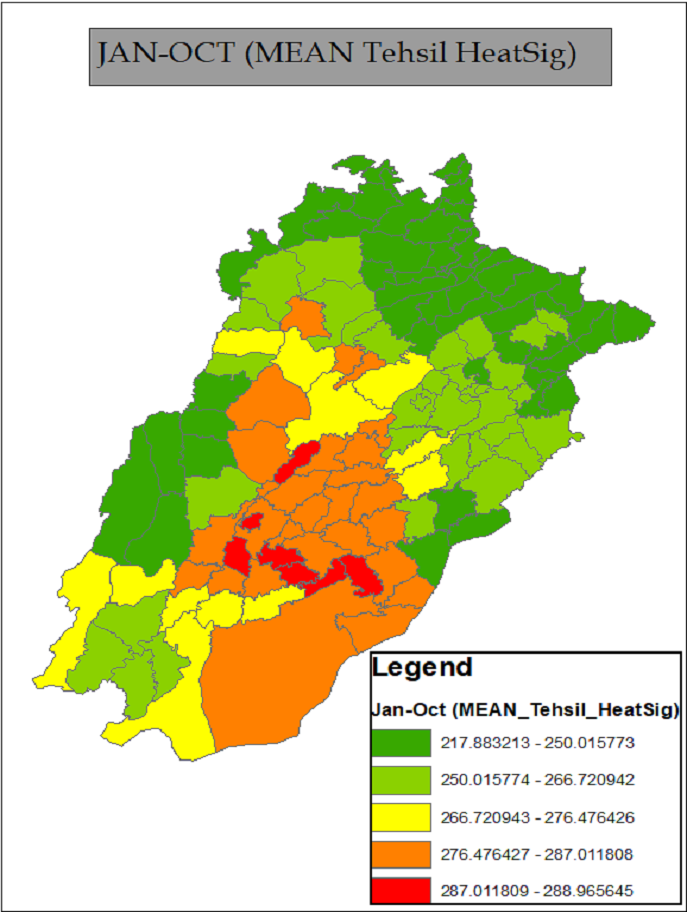} &

			\includegraphics[width=.3\columnwidth]{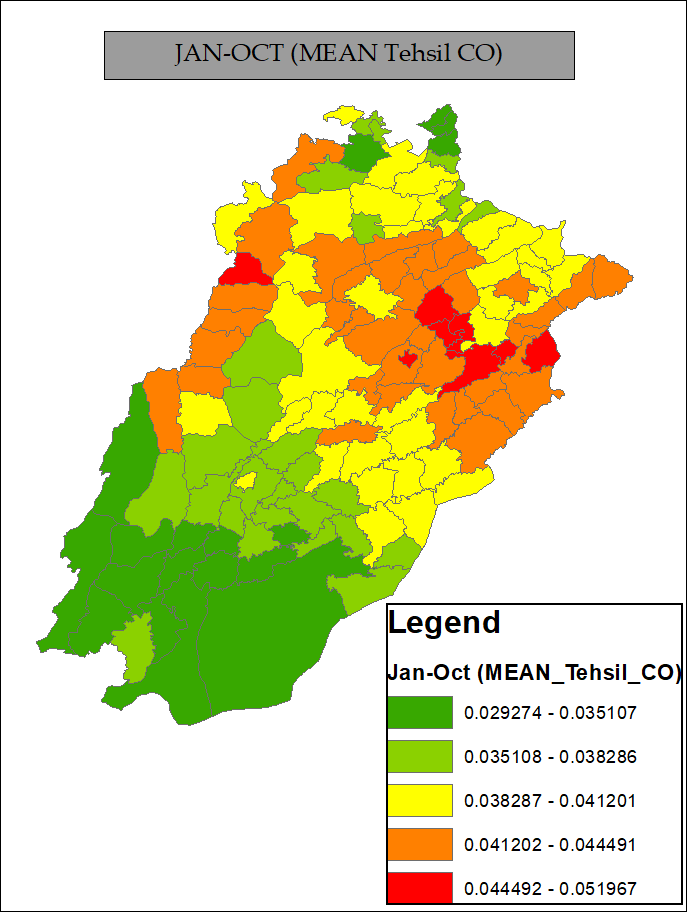} & 
			
			\includegraphics[width=.3\columnwidth] {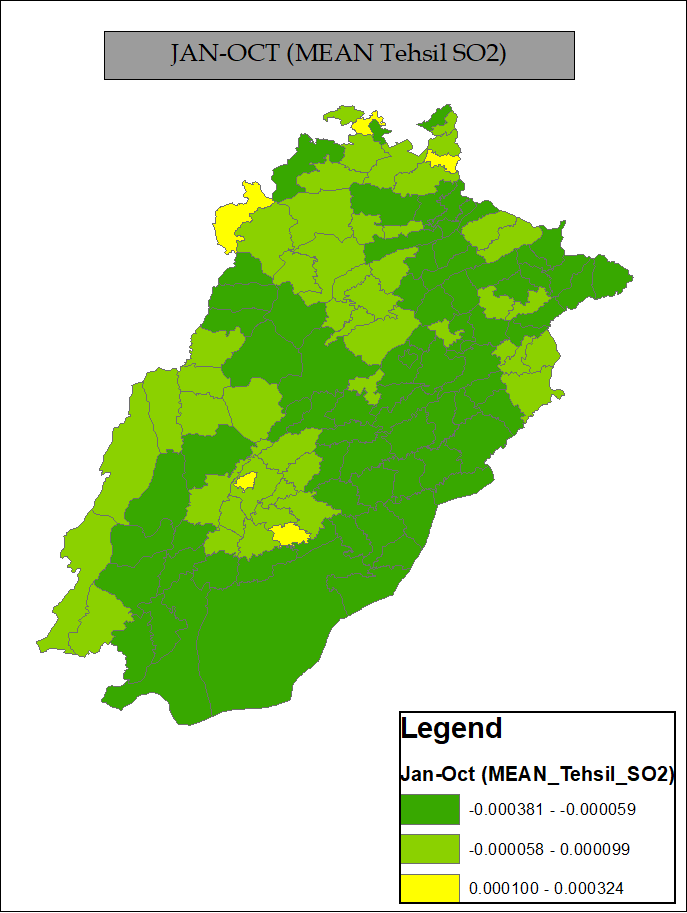} \\ 
			(a) Heat signature  & (b) $CO$  & (c) $SO_2$ \\
			
			\end{tabular}
	\caption{In January-October mean values of Heat Signature, $CO$  and $SO_2$ at Tehsils of Punjab, Pakistan.}
	\label{spectralproperties1}
\end{figure}

\begin{figure}[h]
	\begin{tabular}{ccc}
			\includegraphics[width=.3\columnwidth]{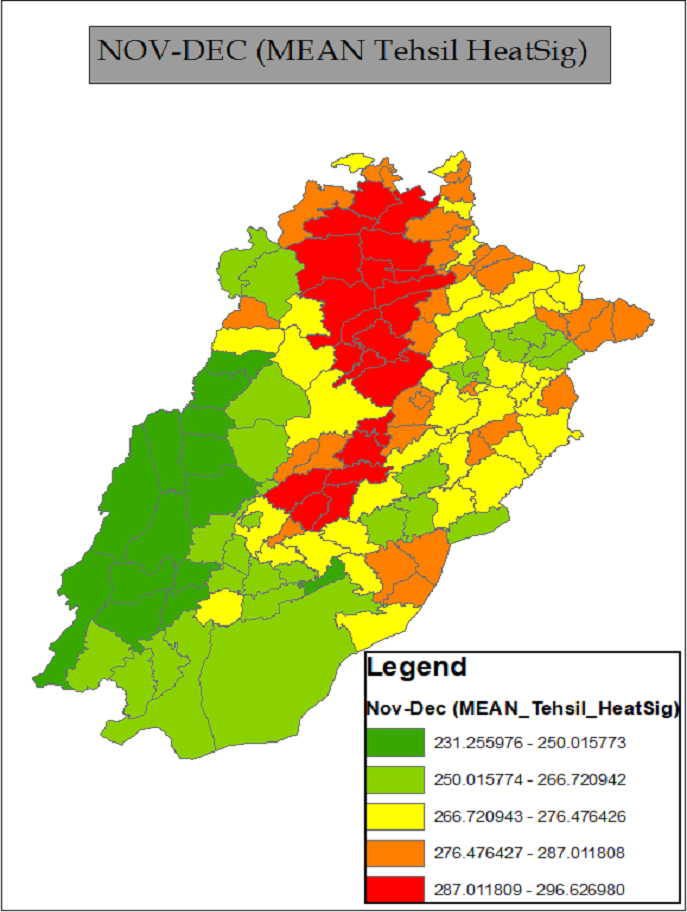} &

			\includegraphics[width=.3\columnwidth]{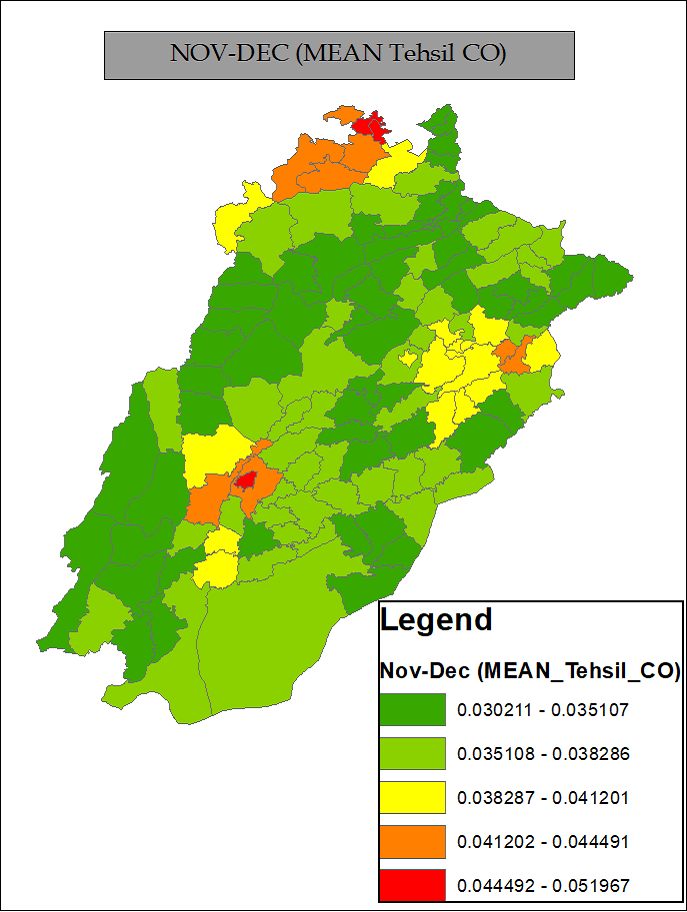} & 
			
			\includegraphics[width=.3\columnwidth] {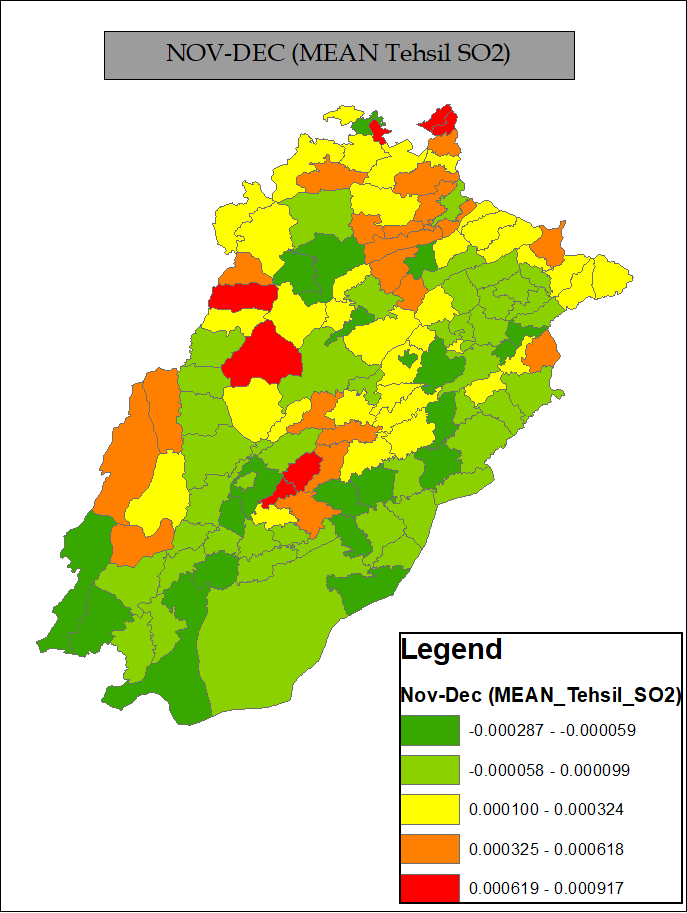} \\ 
			(a) Heat signature  & (b) $CO$  & (c) $SO_2$ \\
			
			\end{tabular}
	\caption{In November-December mean values of Heat Signature, $CO$  and $SO_2$ at Tehsils of Punjab, Pakistan.}
	\label{spectralproperties2}
\end{figure}

\section{Related Work}

Satellite information have been utilized for the tracing of chemical plumes, assist air quality projections, give indication during extraordinary air contamination occasions, gauge model executions, evaluate toxin emissions and study long haul air contamination patterns \cite{dey2022air, dressel2022daily, sutlieff2021using, lavelle2022radiological}. However significant difficulties have been faced in order to overcome this problem. In \cite{inbook}, it was exhibited how Malaysian administrations has already proposed the establishment of a network called Malaysian Remote Sensing Agency (MRSA) for this purpose which is a huge step taken. However, the study area of this research was taken in Penang Island which is the second small-scaled state of Malaysia. Explaining the causes and effects of Aerosol which is a suspension of fine solid and liquid droplets in air which can be natural or anthropogenic. Moreover, the centre of this research was to examine the presentation of their advanced algorithm for delineate PM10 using Landsat 5 having a Thematic Mapper (TM) along with a Multispectral Scanner System (MSS) contrivance. The proposed algorithm models in this study area was demonstrated to be authentic and reliable for corresponding environmental study. 

For acquiring preventative measures it is necessary to have information from time to time about the changes that take place in air pollution degrees. Reflecting this view, there has been an attempt made in \cite{air_pol} for developing a model that will be helpful in analyzing the quality of air using a remotely sensed data which will be easy and quick in process. This study includes vegetation indices, pixel values as well as some urbanization indexes which were used with the help of Landsat ETM+ for evolving regression dependent models with API (Air Pollution Index), this information was further calculated with in-situ air pollutants. This model of multi-variate regression amid Landsat along with the most corresponded variables that gave precise and accurate air pollution image. Almost 90.5\% of exactness was acquired through the comparison that was made between Air Pollution Index Interpolated Images and API modelled.

Brick Kilns have been identified as a major source of pollution, mainly because of poor technology and extremely poor quality of fuel that is used. After China, India is reported to be the second largest producer of bricks. \cite{RAJARATHNAM2014549} shows how over 24 million tons of coal is used in the process per year. In 2017, India’s CPC board ordered all brick kilns to switch to a new method for the production of bricks which is a Zig-Zag kiln. It is shown in ~\cite{rauf2022prospects}, how Zig-Zag kiln used less coal and its structure helps in reducing the fine particulate matter emissions by 51\%. Thus, \cite{rauf2022prospects} also outlays how the tunnel kiln strings better in connection with environmental aspects and quality of brick manufactured. However the drawback is the low investment in this manufacture method and because the process has a requirement of electricity for operating it is a downside as in many parts of India consistent electricity supply is not available.

While the rapid increase in urbanization and development, the demand of brick production is also rising. There have been clear and unfavorable wellbeing results in population living in the prompt region of brick kilns. With regards to \cite{integrated}, which is introducing checking results of every day block furnace stack discharges and inferred variables of it. The study area of this study is the province of Vietnam. Air pollution checking were made on an hourly premise which demonstrated surrounding centralization of $CO$, $SO_2$, $PM2.5$ and $PM10$. Whereas $SO_2$ was observed and resulted of having a high frequency level in comparison with others.  

Air contamination keeps on pulling in increasingly more open consideration. Space-based infrared sensors give a measure to screen air quality in enormous regions. In \cite{he_optimal}, a band choice technique of space based infrared sensors is suggested for urban air contamination discovery, in which perception geometry, ground and climate radiant qualities, and sensor system framework commotions are coordinated. This proposed system is executed to investigate an ideal band for identifying four unique sorts of vaporous poisons and segregating aerosol particle contamination to demonstrate its value. This material science based investigation strategy can be utilized in space-based infrared sensor structure, particularly for obtaining accessible information in districts with generally low pollutant fixations. This methodology can likewise be utilized to recover pollutant concentrations.

 \begin{figure}[h]
 \centering
	\begin{tabular}{cc}
			\includegraphics[width=.3\columnwidth]{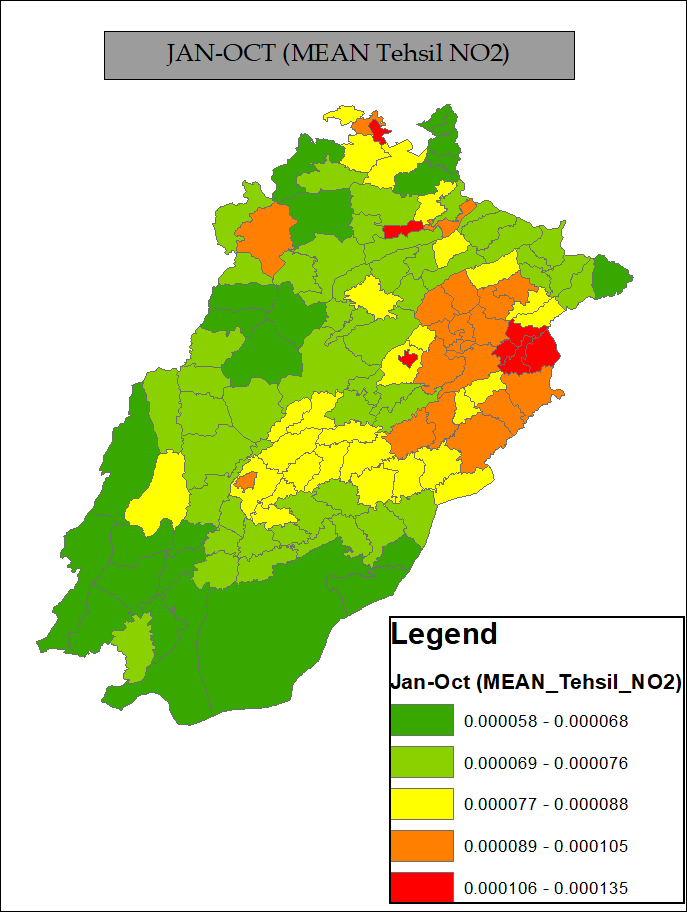} &

			\includegraphics[width=.3\columnwidth]{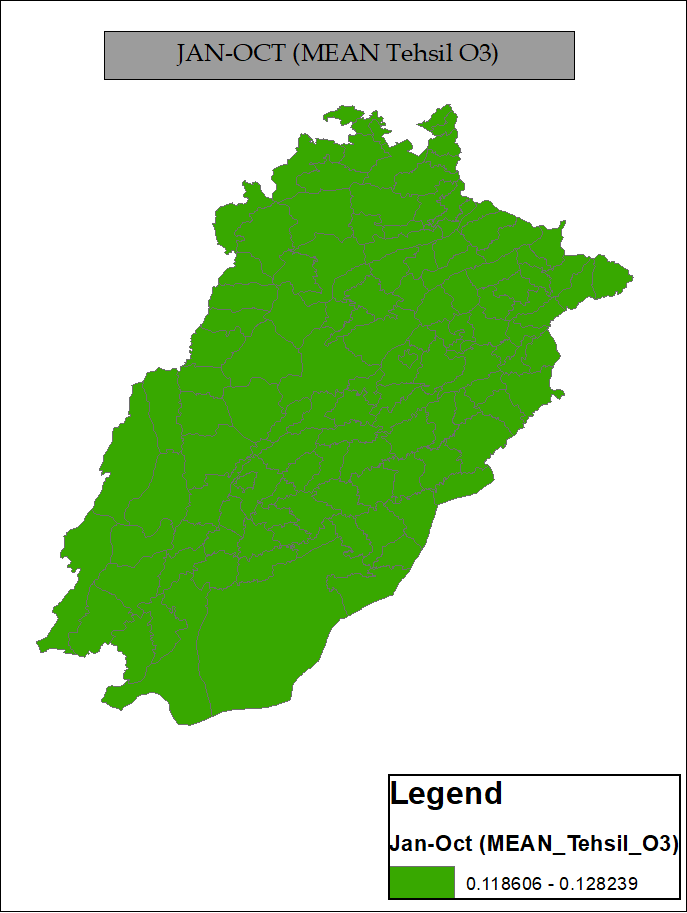} \\
			
			(a) $NO_2$  & (b) $O_3$  \\
			
			\end{tabular}
	\caption{In January-October mean values of $NO_2$  and $O_3$ at Tehsils of Punjab, Pakistan.}
	\label{spectralproperties3}
 \end{figure}

  \begin{figure}[h]
  \centering
	\begin{tabular}{cc}
			\includegraphics[width=.3\columnwidth]{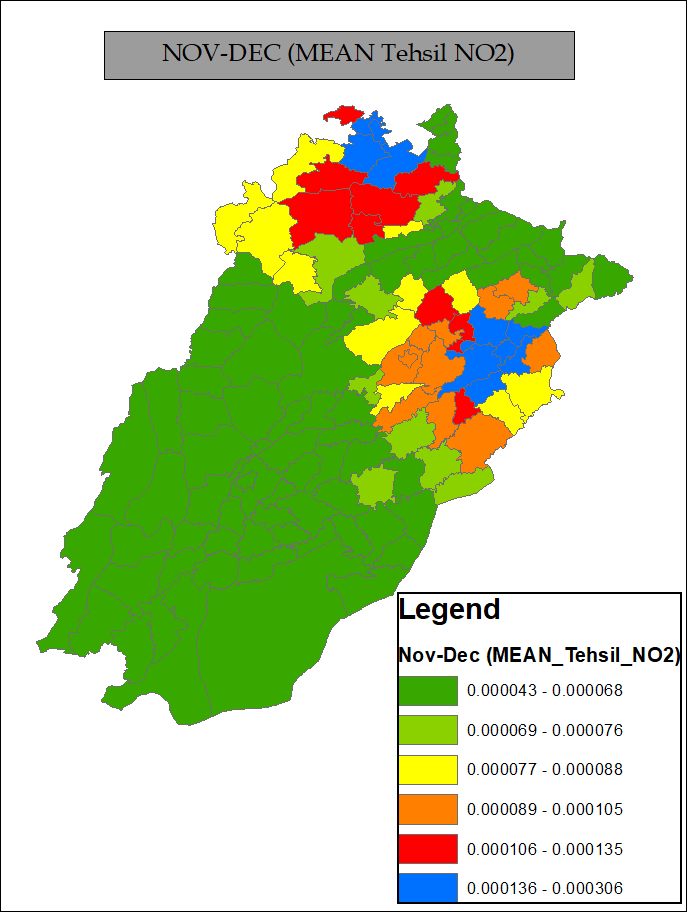} &

			\includegraphics[width=.3\columnwidth]{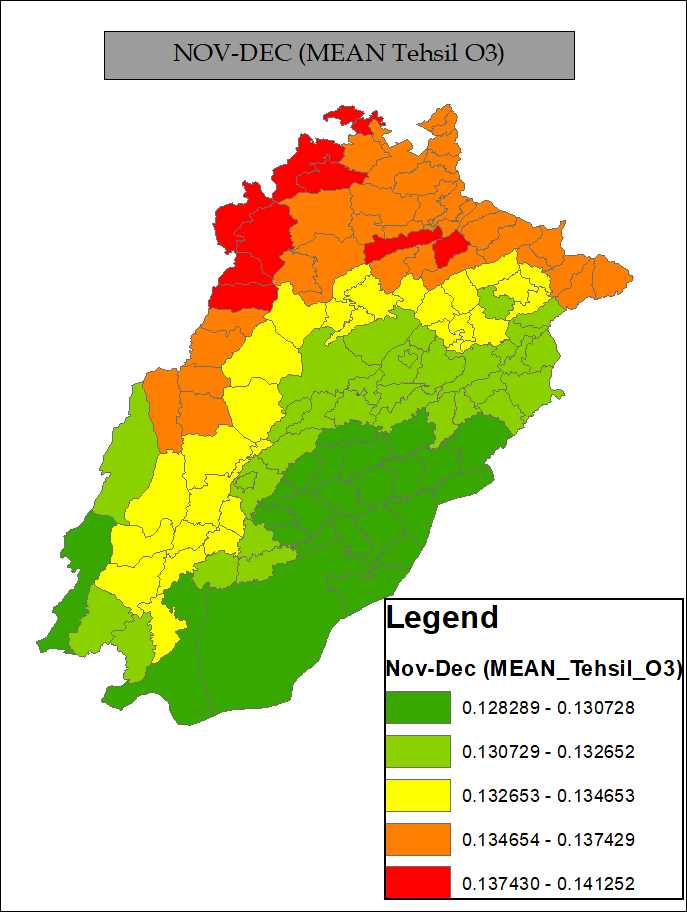} \\
			
			(a) $NO_2$  & (b) $O_3$  \\
			
			\end{tabular}
	\caption{In November-December mean values of $NO_2$  and $O_3$ at Tehsils of Punjab, Pakistan.}
	\label{spectralproperties4}
\end{figure}


\section{Proposed Method}
To detect the environmental impact caused by operation of brick kilns during SMOG period is a challenging problem for which theoretical background of remote sensing data such as, heat signature and gaseous emissions for the detection of heat emission and air pollution. RGB bands does not give us the information about heat signature and gaseous emissions. RGB bands only tells us about the kilns locations, we can see whether the kiln is present on the specified image or not. What we need is the heat signature and data on gaseous emissions to know if the kiln is working or not which can be calculated using Landsat-8 satellite's Thermal Infrared 1 Band (band 10) and Sentinel-5 satellite's TROPOMI (TROPOspheric Monitoring Instrument). 

The image resolution for which we are getting data from Landsat-8 satellite's Thermal Infrared 1 Band (band 10) is accurate for heat signature values i.e. 30 meters. The data we are getting from Sentinel-5 satellite for gaseos emissions: $CO$, $NO_2$, $SO_2$ and $O_3$ uses the patch of $7 \times 7$ $km^2$ which is not much accurate for kilns because a brick kiln is usually $30$ meters wide. So we are fusing two different types of data to get close to our result which is detecting illegal operation of kilns [see Fig.~\ref{fig:fusion}]. Current data and state of the art techniques cannot accurately find the kiln heat signature and concentration of gases emitted by kilns. This is why we are looking at tehsils level to measure these parameters before and during SMOG period, so the Environment Protection Department (EPD) can be notified that these tehsils are violating the protocols set by government.

The heat signature and gaseos emissions dataset, we generated in this paper, provides the basis for the detection of illegal operation of kilns during SMOG period (see Fig. \ref{fig:GroundTruth1}). The proposed architecture we aim to implement is shown in Fig. \ref{fig:fusion}.


\subsection{Dataset}
 The main source of data set is Punjab Brick Kiln Census\footnote{http://dashboards.urbanunit.gov.pk/brick\_kiln\_dashboard/} and Kiln-Net paper~\cite{nazir2020kiln}. The data was retrieved for 9348 brick kiln locations. The location coordinates were separated from the data. We used those coordinates to generate the data of heat signature, $CO$, $NO_2$, $SO_2$ and $O_3$ using Google Earth Engine (GEE) which is a remote sensing tool. We calculated mean, median and max of all these parameters along two time periods. The time periods are: January- October, 2019 and November- December, 2019. The reason for selection of these time period is the ground truth data: No. of FIR lodged against brick kilns and the number of brick kilns sealed in SMOG period in year 2019 (see Fig.~\ref{fig:GroundTruth}).

As the area of interest is Punjab, Pakistan. We extracted shape files of tehsils of Punjab using ArcGIS Online to gather the data on tehsil level using GEE. 

\subsection{Heat signature and gaseous emissions}


Landsat-8 satellite images have additional bands with different wavelengths. These bands represent different type of information which cannot be perceived by human eye. The satellite senses energy (light from the sun) reflecting from the earth’s surface back to the sensor at different wavelengths depending on the composition of the object the light struck. For example, if we take bands 5, 6, 7 and stack those in the RGB color space so that our screens can display infrared and near-infrared light. Similarly, Thermal Infrared 1 Band (band 10) is used in proposed methodology to determine temperature or heat signature values of earth's surface. We can use different combination of these bands to find different properties of earth such as vegetation index, moisture index, built-up index etc. The data we get from these bands uses image resolution of 15-30 meters per pixel.


Satellite imagery also provide information about the gaseous emission in the atmosphere. We used Sentinel-5 satellite to get information about concentration of different gases such as $CO$, $NO_2$, $SO_2$, $O_3$, $HCHO$, $CH_4$ and aerosol optical depth. The spectrometer covers the spectral range from the ultraviolet to the shortwave infrared (270-2385 nm) with a spatial resolution of 7 x 7 $km^2$. 


 \begin{figure}
	\begin{tabular}{ccc}
			\includegraphics[width=.3\columnwidth]{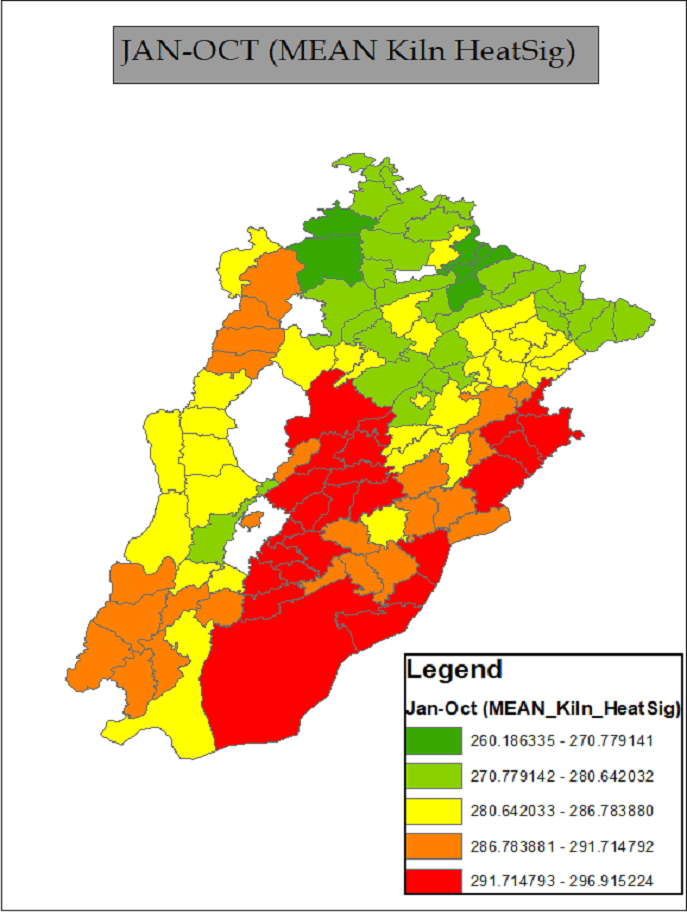} &

			\includegraphics[width=.3\columnwidth]{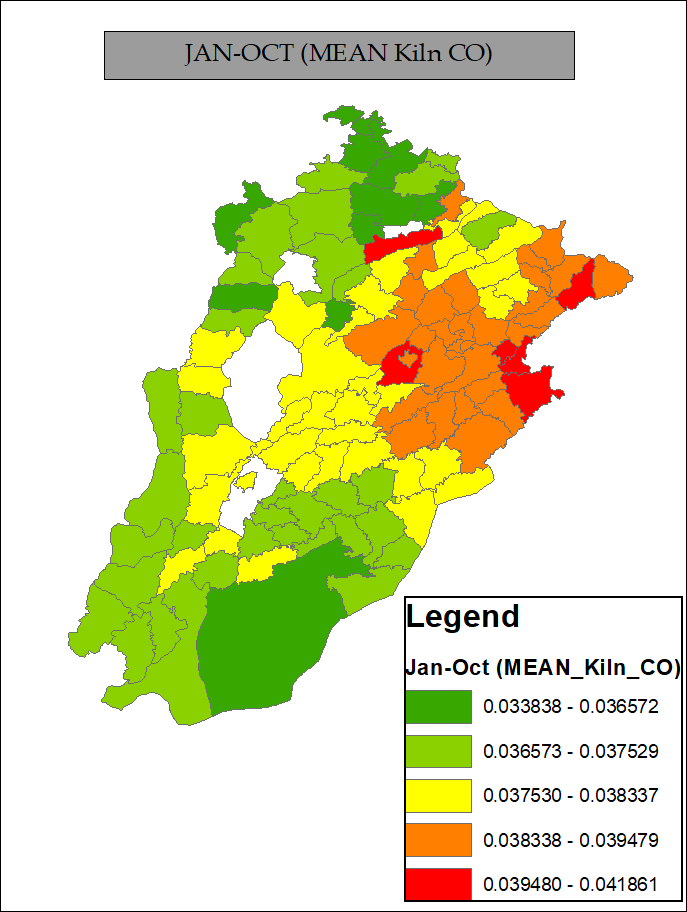} & 
			
			\includegraphics[width=.3\columnwidth] {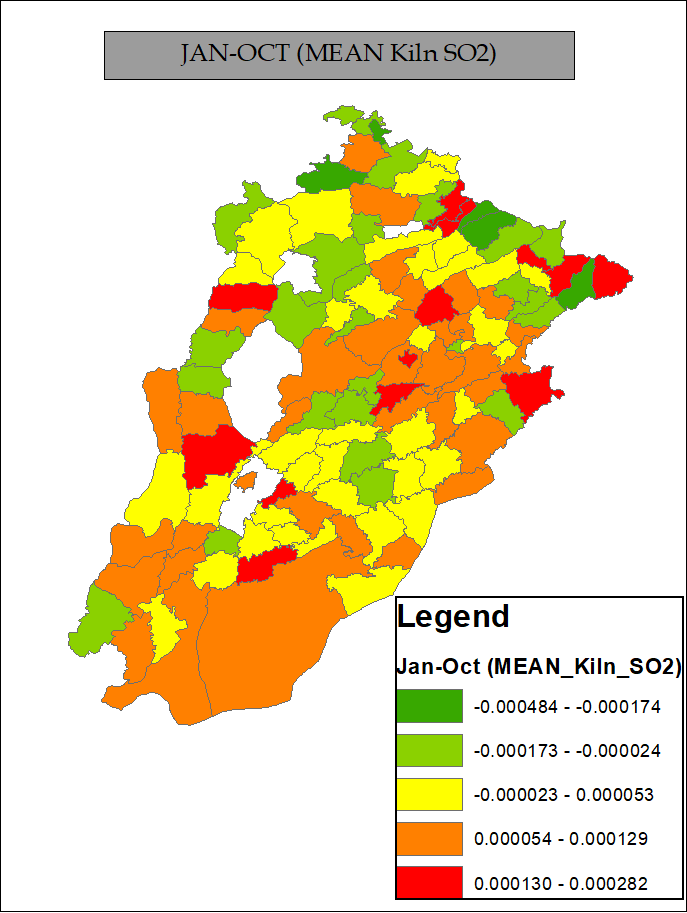} \\ 
			(a) Heat signature  & (b) $CO$  & (c) $SO_2$ \\
			
			\end{tabular}
	\caption{In January-October mean values of Heat Signature, $CO$  and $SO_2$ at Kiln locations of Punjab, Pakistan.}
	\label{spectralproperties5}

	\begin{tabular}{ccc}
			\includegraphics[width=.3\columnwidth]{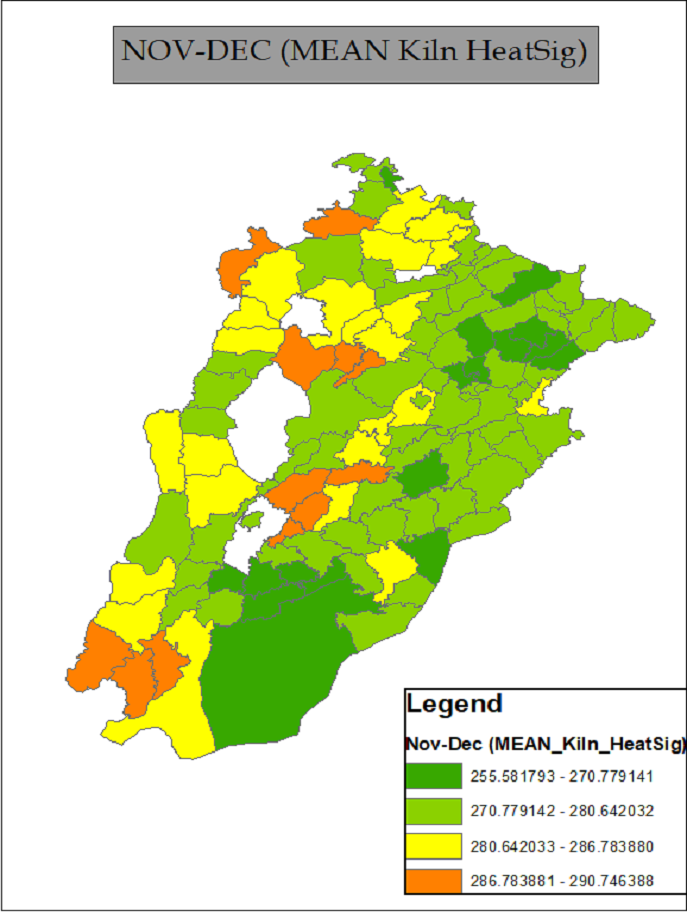} &

			\includegraphics[width=.3\columnwidth]{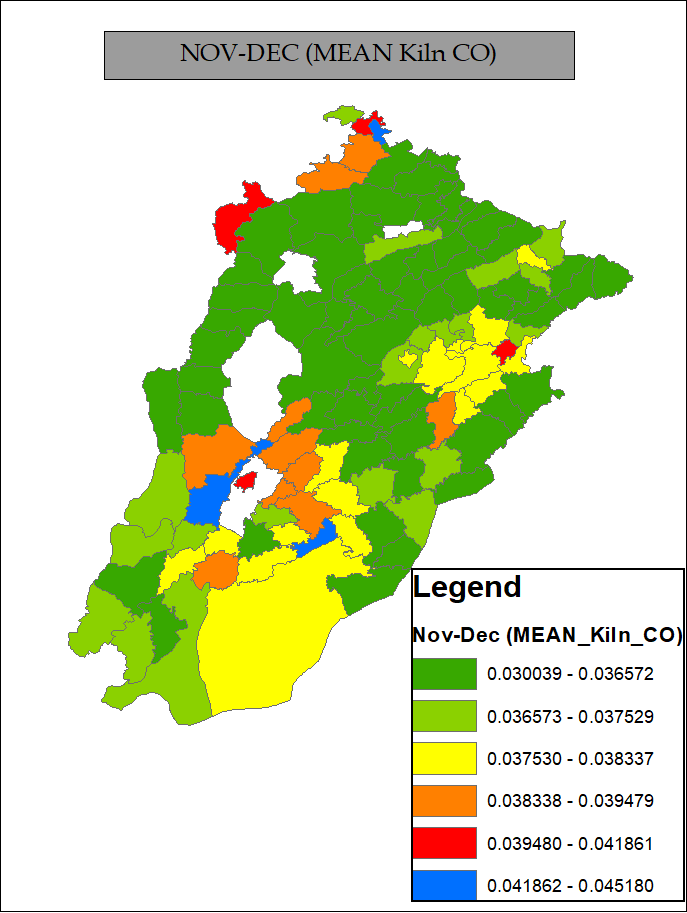} & 
			
			\includegraphics[width=.3\columnwidth] {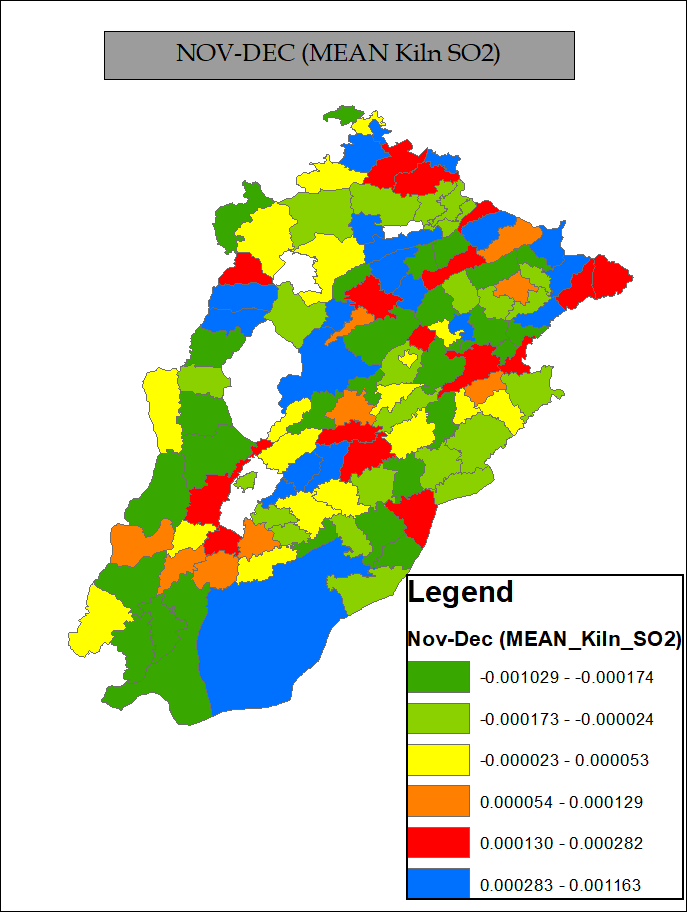} \\ 
			(a) Heat signature  & (b) $CO$  & (c) $SO_2$ \\
			
			\end{tabular}
	\caption{In November-December mean values of Heat Signature, $CO$  and $SO_2$ at Kiln locations of Punjab, Pakistan.}
	\label{spectralproperties6}
\end{figure}

\section{Experimental Results}
In this section we discussed the comparison of different illustrations on data set and the results we obtained from our proposed method. Discussion on obtained results and its comparison is also mentioned in this section. 

\begin{figure}[h]
\centering
	\begin{tabular}{cc}
			\includegraphics[width=.3\columnwidth]{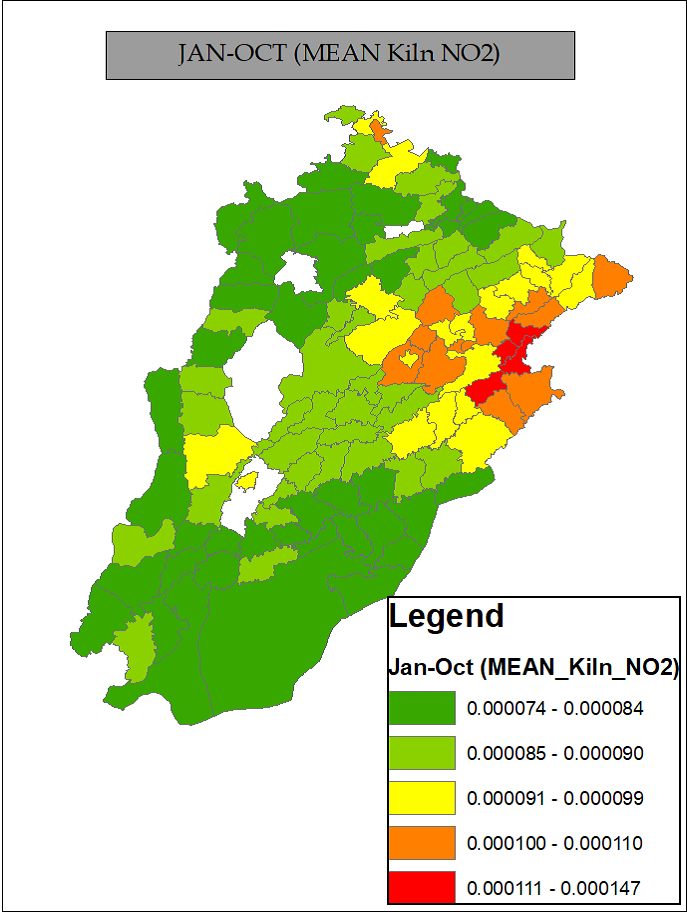} &

			\includegraphics[width=.3\columnwidth]{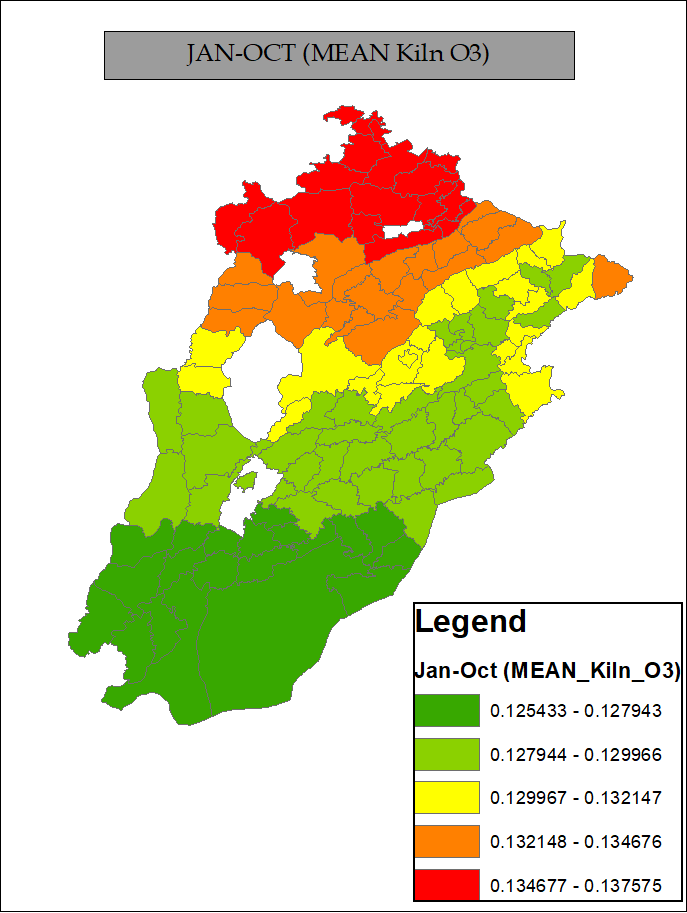} \\
			
			(a) $NO_2$  & (b) $O_3$  \\
			
			\end{tabular}
	\caption{In January-October mean values of $NO_2$  and $O_3$ at Kiln locations of Punjab, Pakistan.}
	\label{spectralproperties7}
 \end{figure}

  \begin{figure}[h]
  \centering
	\begin{tabular}{cc}
			\includegraphics[width=.3\columnwidth]{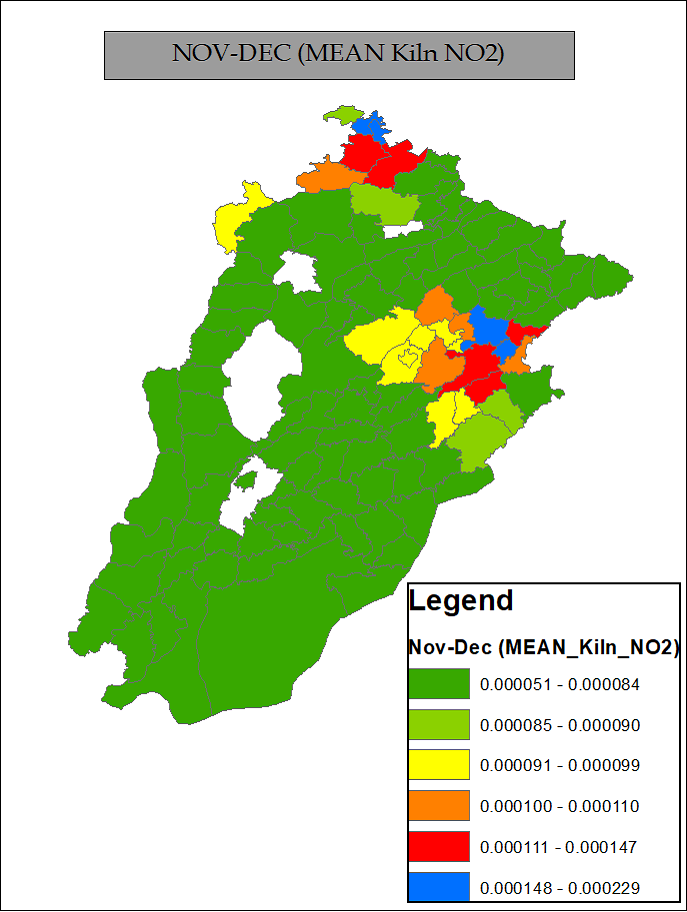} &

			\includegraphics[width=.3\columnwidth]{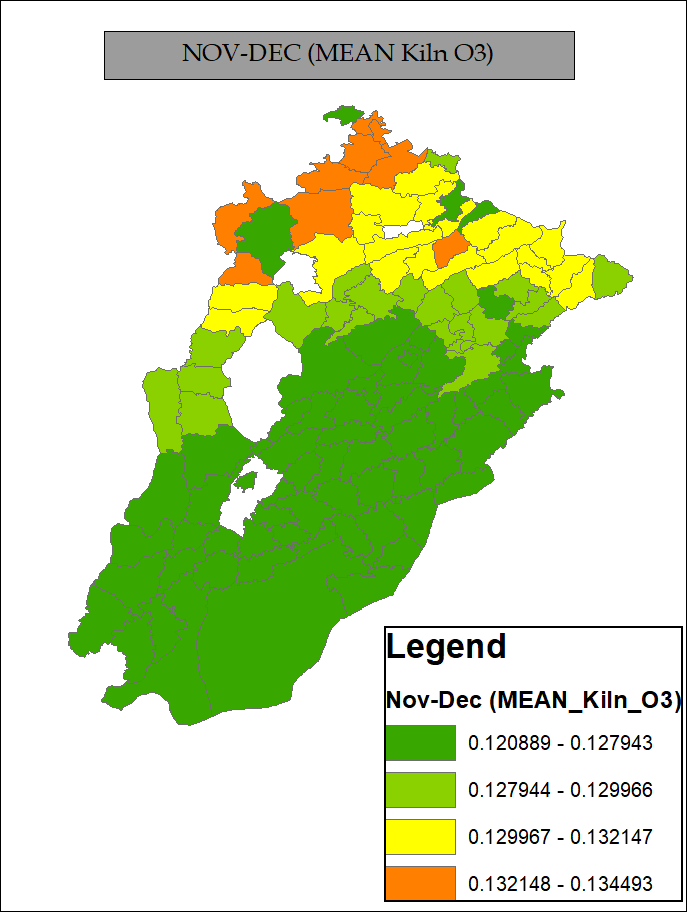} \\
			
			(a) $NO_2$  & (b) $O_3$  \\
			
			\end{tabular}
	\caption{In November-December mean values of $NO_2$  and $O_3$ at Kiln locations of Punjab, Pakistan.}
	\label{spectralproperties8}
\end{figure}
\subsection{Tehsil Level Analysis}
In this section we discuss the illustrations of tehsils, for which we get the results of mean values on heat signature, $CO$, $SO_2$, $NO_2$ and $O_3$ during January-October and November-December 2019. 

The heat signature for Jan-Oct period (Fig. \ref{spectralproperties1}a) in northern Punjab is comparatively lower than heat signature for Nov-Dec period (Fig. \ref{spectralproperties2}a). This does not necessarily mean that kilns activity has increased in later period. Similarly, heat signature for Jan-Oct period (Fig. \ref{spectralproperties1}a) in southern Punjab is comparatively higher than heat signature for Nov-Dec period (Fig. \ref{spectralproperties2}a). Heat Signature on tehsil level depends on other factors too e.g. industrial activities. So we cannot say anything for sure about kilns activity by looking only at heat signature on tehsil level.

The $CO$ emission for Jan-Oct period (Fig. \ref{spectralproperties1}b) in northern Punjab is comparatively higher than $CO$ emission for Nov-Dec period (Fig. \ref{spectralproperties2}b). This does not definitely mean if the kilns activity has decreased in later period. Similarly, $CO$ emission for Jan-Oct period (Fig. \ref{spectralproperties1}b) in southern Punjab is almost same as $CO$ emission for Nov-Dec period (Fig. \ref{spectralproperties2}b). $CO$ emission on tehsil level depends on other factors too e.g. burning of oils, gases, coal, wood and fires. We cannot say anything about kilns activity by looking only at $CO$ emission on tehsil level.

The $SO_2$ emission for Jan-Oct period (Fig. \ref{spectralproperties1}c) in Punjab is comparatively lower than $SO_2$ emission for Nov-Dec period (Fig. \ref{spectralproperties2}c). This does not mean that kilns activity has increased in later period because these results show us the overall activity of that tehsil. Burning of fossils fuels such as coal, oil, diesel etc. are the main cause behind emission of $SO_2$. Kiln activity cannot be determined by looking at $SO_2$ emission on tehsil level.

The $NO_2$ emission for Jan-Oct period (Fig. \ref{spectralproperties3}a) in southern Punjab is comparatively higher than $NO_2$ emission for Nov-Dec period (Fig. \ref{spectralproperties4}a). This does not certainly mean that kilns activity has decreased in later period. Similarly, $NO_2$ emission for Nov-Dec period (Fig. \ref{spectralproperties4}a) in northern Punjab has somewhat increased as compared to Jan-Oct period (Fig. \ref{spectralproperties3}a). $NO_2$ emission on tehsil level depends on other factors such as burning of fuels from cars, trucks, buses, power plants, brick kilns etc. 

The $O_3$ emission for Jan-Oct period (Fig. \ref{spectralproperties3}b) in all tehsils of Punjab is lower than $O_3$ emission for Nov-Dec period (Fig. \ref{spectralproperties4}b). This does not depicts that kilns activity has increased in later period. Some chemicals that react to form ozone $O_3$ are oil refining, aviation, petrochemicals, motor vehicle exhaust, bushfires and burning off. 70\% of the nitrogen oxides and 50\% of the organic chemicals that form ozone $O_3$ are produced by motor vehicle exhaust. Organic chemicals emission because of vegetation can also help form ozone $O_3$.
\subsection{Kiln Level Analysis}
In this section we discuss the illustrations of kilns, for which we get the results of mean values on heat signature, $CO$, $SO_2$, $NO_2$ and $O_3$ during Jan-Oct and Nov-Dec 2019. 

The heat signature for Jan-Oct period (Fig. \ref{spectralproperties5}a) in Punjab is comparatively higher than heat signature for Nov-Dec period (Fig. \ref{spectralproperties6}a). This means that kilns activity has significantly decreased in later period. Out of 128 tehsils, we found that heat signature of 27 tehsils was higher in Nov-Dec period than Jan-Oct period. This means that the operation of kilns has increased in those tehsils during SMOG period.

The $CO$ emission for Jan-Oct period (Fig. \ref{spectralproperties5}b) in Punjab is comparatively higher than $CO$ emission for Nov-Dec period (Fig. \ref{spectralproperties6}b). This shows that kilns activity has notably decreased during the SMOG period due to which $CO$ emissions are less because of the kilns activity. For the areas in which $CO$ emission is higher during Nov-Dec can be because of illegal kiln activity.

The $SO_2$ emission for Jan-Oct period (Fig. \ref{spectralproperties5}c) in Punjab is comparatively lower than $SO_2$ emission for Nov-Dec period (Fig. \ref{spectralproperties6}c). This can be because of the kilns activity has increased in later period or it can be due to burning of fossils fuels such as coal, oil, diesel etc. are the main cause behind emission of $SO_2$.

The $NO_2$ emission for Jan-Oct period (Fig. \ref{spectralproperties7}a) in Punjab is higher than $NO_2$ emission for Nov-Dec period (Fig. \ref{spectralproperties8}a). The diagram show us that kilns activity has significantly decreased in Nov-Dec period. In some tehsils, we see the increase in $NO_2$ emission but that can be because of kiln activity or burning of fuels from cars, trucks, buses, power plants etc.

The $O_3$ emission for Jan-Oct period (Fig. \ref{spectralproperties7}b) in all tehsils of Punjab is higher than $O_3$ emission for Nov-Dec period (Fig. \ref{spectralproperties8}b). The data depicts in diagram that kilns activity has decreased during SMOG period which is Nov-Dec. We can see the red and orange spikes in Jan-Oct period (Fig. \ref{spectralproperties7}b) are turning to orange and yellow spikes showing us that emissions of $O_3$ are highly decreasing in Nov-Dec period.

\subsection{Qualitative Evaluation}

All the factors discussed above in both sections are directly related to kiln activity and SMOG. What we need to understand is brick kiln activity has an immediate connection to SMOG. The smoke coming out of brick kilns produce several toxic gases such as $CO$, $SO_2$, $NO_2$ and $O_3$. 

The diagrams of tehsils portray compelling results but those diagrams indicate the results of whole tehsil instead of just the brick kilns, due to which we do not have absolute results for brick kilns activity by examining results of only tehsils. Owing to this reason we decided to go deeper and calculated the data of heat signature, $CO$, $SO_2$, $NO_2$ and $O_3$ on kilns level.

The new data represented more accurate results and we can illustrate from those new diagrams that brick kilns activity is tremendously decreasing during the SMOG period because of Punjab government's restriction but there are still some tehsils in which authorities are not monitoring the situation and not working properly. We can see from the heat signature, $CO$, $NO_2$ and $O_3$ emission of kilns that operations of kilns are decreased during Nov-Dec period. However, we get conflicting result from $SO_2$ diagrams which tells us that $SO_2$ emissions are increasing during SMOG period but that can be because of other factors such as burning of fossil fuels.

\begin{figure}[h]
\centering
\begin{tabular}{cc}
			\includegraphics[width=.35\columnwidth]{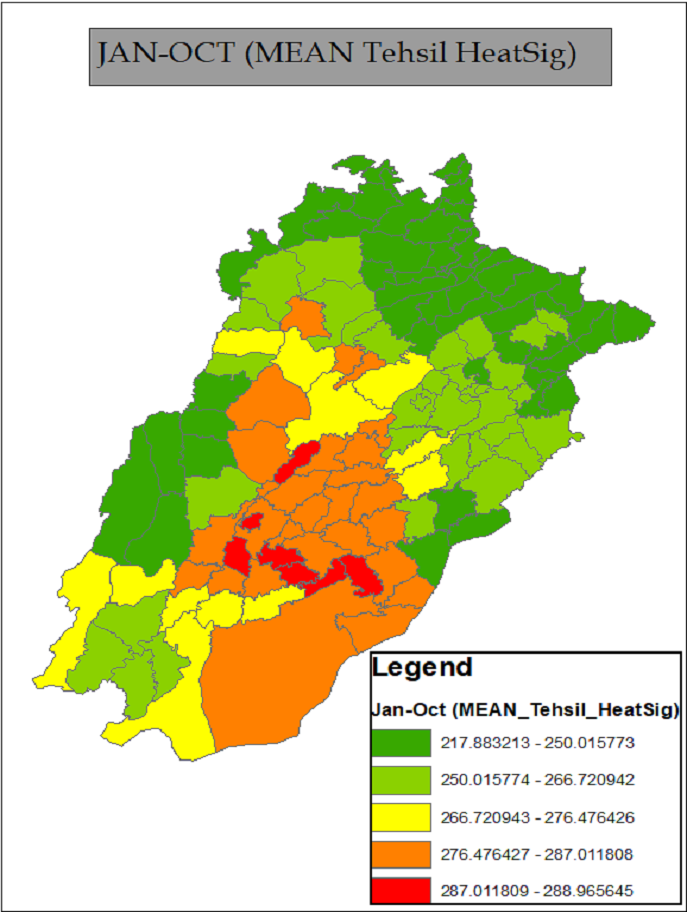} &

			\includegraphics[width=.35\columnwidth]{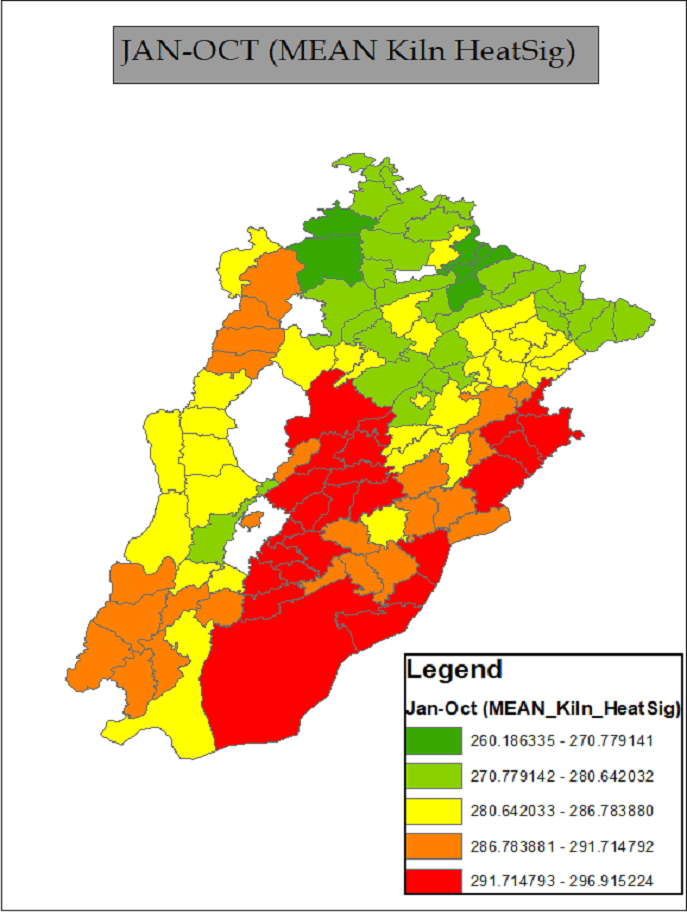} \\
			
			(a) Teshils  & (b) Kilns  \\
			
			\end{tabular}
	\caption{In January-October mean values of heat signature at Tehsils and Kiln locations of Punjab, Pakistan.}
	\label{spectralproperties9}
\end{figure}
\begin{figure}[h]
\centering
    \begin{tabular}{cc}
			\includegraphics[width=.35\columnwidth]{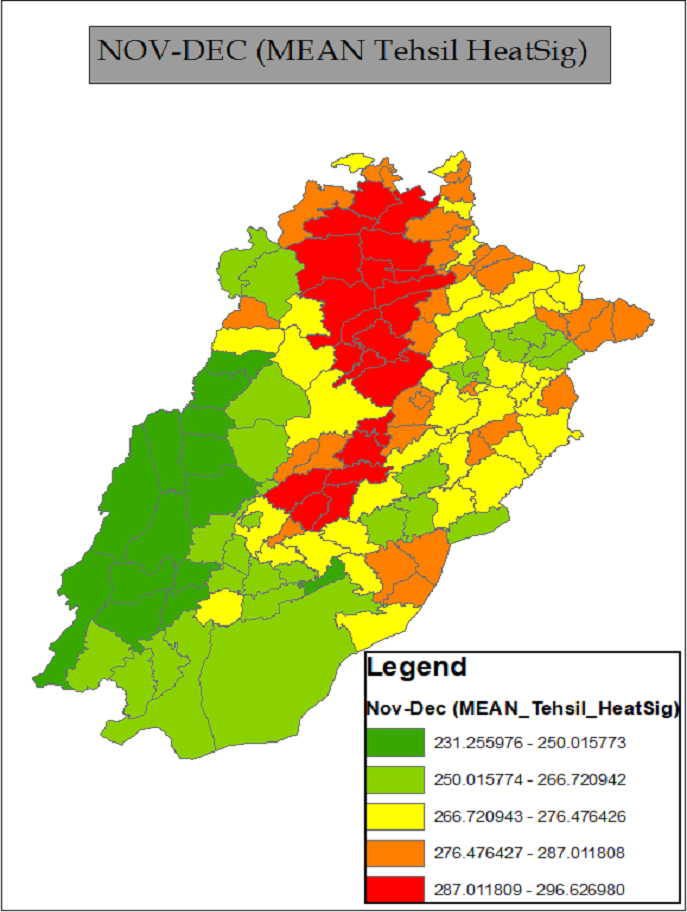} &

			\includegraphics[width=.35\columnwidth]{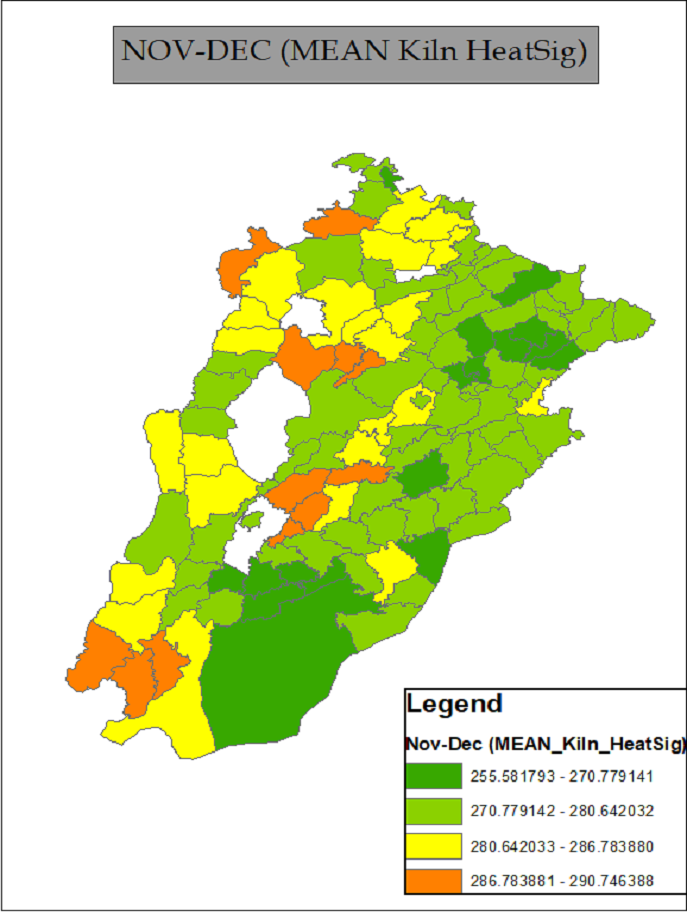} \\
			
			(a) Teshils  & (b) Kilns  \\
			
			\end{tabular}
	\caption{In November-December mean values of heat signature at Tehsils and Kiln locations of Punjab, Pakistan.}
	\label{spectralproperties10}
\end{figure}

Looking at Fig. \ref{spectralproperties9}a and \ref{spectralproperties9}b, we can clearly see from the diagram, mean heat signature of kilns are higher as compared to tehsils because data of tehsils incorporate the mean data of the whole tehsils whereas kiln data is only specific to the kilns of those tehsils. Similarly, from Fig. \ref{spectralproperties10}a and \ref{spectralproperties10}b, we see that mean heat signature of kilns are getting lower as compare to tehsils as kiln data is only focusing on kilns of tehsils as compared to whole activity of the tehsils. From Fig. \ref{spectralproperties9}b and \ref{spectralproperties10}b, we can extract that kiln activity is decreasing during winter SMOG period and there are some tehsils in which activity is increasing and are violating the government's restrictions.
\section{Conclusion}
For detection of illegal kiln activity, our proposed method found from the remote sensing data that around 27 tehsils (in Punjab, Pakistan) out of 128 were causing the increase in SMOG. Most of the brick kilns burn coal as a fuel source, resulting in SO2 and particulate matter (PM) emissions, causing poor air quality and associated health problems. For future work, we can calculate the concentration of greenhouse gases emitted or absorbed by trees to identify better results on tehsil level. To increase the impact of the overall work we can work together with International Labor Organization (ILO) and National Institute of Health (NIH) to come up with actionable items and policy briefs to improve the human rights condition of the labor workforce. We can also work with Environment Protection Department (EPD), Pakistan to enforce the preventive measures on tehsil level regarding illegal kiln operation.
\section{Acknowledgement}
We acknowledge the partial financial support of National Center for Robotics and Automation (NCRA), Pakistan and National Agriculture Robotics Laboratory (NARL) at LUMS in carrying out this research.

\bibliographystyle{IEEEtran}
\bibliography{refs}

\newcommand{\noopsort}[1]{} \newcommand{\printfirst}[2]{#1}
  \newcommand{\singleletter}[1]{#1} \newcommand{\switchargs}[2]{#2#1}
\begin{thebibliography}{10}
\providecommand{\url}[1]{#1}
\csname url@samestyle\endcsname
\providecommand{\newblock}{\relax}
\providecommand{\bibinfo}[2]{#2}
\providecommand{\BIBentrySTDinterwordspacing}{\spaceskip=0pt\relax}
\providecommand{\BIBentryALTinterwordstretchfactor}{4}
\providecommand{\BIBentryALTinterwordspacing}{\spaceskip=\fontdimen2\font plus
\BIBentryALTinterwordstretchfactor\fontdimen3\font minus
  \fontdimen4\font\relax}
\providecommand{\BIBforeignlanguage}[2]{{%
\expandafter\ifx\csname l@#1\endcsname\relax
\typeout{** WARNING: IEEEtran.bst: No hyphenation pattern has been}%
\typeout{** loaded for the language `#1'. Using the pattern for}%
\typeout{** the default language instead.}%
\else
\language=\csname l@#1\endcsname
\fi
#2}}
\providecommand{\BIBdecl}{\relax}
\BIBdecl

\bibitem{haque2022impact}
S.~E. Haque, M.~M. Shahriar, N.~Nahar, and M.~S. Haque, ``Impact of brick kiln
  emissions on soil quality: A case study of ashulia brick kiln cluster,
  bangladesh,'' \emph{Environmental Challenges}, vol.~9, p. 100640, 2022.

\bibitem{hussain2022brick}
A.~Hussain, N.~U. Khan, M.~Ullah, M.~Imran, M.~Ibrahim, J.~Hussain, H.~Ullah,
  I.~Ullah, I.~Ahmad, M.~U. Khan \emph{et~al.}, ``Brick kilns air pollution and
  its impact on the peshawar city,'' \emph{Pollution}, vol.~8, no.~4, pp.
  1266--1273, 2022.

\bibitem{subhanullah2022assessment}
M.~Subhanullah, S.~Ullah, M.~F. Javed, R.~Ullah, T.~A. Akbar, W.~Ullah, S.~A.
  Baig, M.~Aziz, A.~Mohamed, and R.~U. Sajjad, ``Assessment and impacts of air
  pollution from brick kilns on public health in northern pakistan,''
  \emph{Atmosphere}, vol.~13, no.~8, p. 1231, 2022.

\bibitem{SmogWiki7online}
``Smog - {Wikipedia},'' \url{https://en.wikipedia.org/wiki/Smog}, (Accessed on
  05/15/2020).

\bibitem{arun2022planning}
S.~Arun, U.~Anand, P.~Bhattacharjee \emph{et~al.}, ``Planning, analysis, and
  design of smog-free tower with louvers in kolkata,'' in \emph{Advances in
  Construction Management}.\hskip 1em plus 0.5em minus 0.4em\relax Springer,
  2022, pp. 3--11.

\bibitem{kiln_causes}
B.~Skinder, A.~Pandit, A.~Qayoom, and B.~Ahmad, ``Brick kilns: Cause of
  atmospheric pollution,'' \emph{Pollution Effects \& Control}, vol.~02, 08
  2014.

\bibitem{integrated}
L.~H.A. and O.~N.T.K., ``Integrated assessment of brick kiln emission impacts
  on air quality,'' \emph{Environment Monitoring and Assessment}, vol. 171, p.
  381–394, 2010.

\bibitem{koukouli2022air}
M.-E. Koukouli, A.~Pseftogkas, D.~Karagkiozidis, I.~Skoulidou, T.~Drosoglou,
  D.~Balis, A.~Bais, D.~Melas, and N.~Hatzianastassiou, ``Air quality in two
  northern greek cities revealed by their tropospheric no2 levels,''
  \emph{Atmosphere}, vol.~13, no.~5, p. 840, 2022.

\bibitem{dey2022air}
S.~Dey and S.~Chowdhury, ``Air quality management in india using satellite
  data,'' in \emph{Asian Atmospheric Pollution}.\hskip 1em plus 0.5em minus
  0.4em\relax Elsevier, 2022, pp. 239--254.

\bibitem{dressel2022daily}
I.~M. Dressel, M.~A.~G. Demetillo, L.~M. Judd, S.~J. Janz, K.~P. Fields,
  K.~Sun, A.~M. Fiore, B.~C. McDonald, and S.~E. Pusede, ``Daily satellite
  observations of nitrogen dioxide air pollution inequality in new york city,
  new york and newark, new jersey: Evaluation and application,''
  \emph{Environmental Science \& Technology}, vol.~56, no.~22, pp.
  15\,298--15\,311, 2022.

\bibitem{sutlieff2021using}
G.~Sutlieff, L.~Berthoud, and M.~Stinchcombe, ``Using satellite data for cbrn
  (chemical, biological, radiological, and nuclear) threat detection,
  monitoring, and modelling,'' \emph{Surveys in Geophysics}, vol.~42, no.~3,
  pp. 727--755, 2021.

\bibitem{lavelle2022radiological}
M.~Lavelle, ``Radiological and nuclear,'' in \emph{CBRNE: Challenges in the
  21st Century}.\hskip 1em plus 0.5em minus 0.4em\relax Springer, 2022, pp.
  79--99.

\bibitem{inbook}
H.~S. Lim, M.~Z. Mat~Jafri, K.~Abdullah, and W.~Jeng, \emph{Air Pollution
  Determination Using Remote Sensing Technique}, 10 2009.

\bibitem{air_pol}
C.~Mozumder, K.~Venkata, and D.~Pratap, ``Air pollution modeling from remotely
  sensed data using regression techniques,'' \emph{Journal of the Indian
  Society of Remote Sensing}, vol.~41, 10 2012.

\bibitem{RAJARATHNAM2014549}
\BIBentryALTinterwordspacing
U.~Rajarathnam, V.~Athalye, S.~Ragavan, S.~Maithel, D.~Lalchandani, S.~Kumar,
  E.~Baum, C.~Weyant, and T.~Bond, ``Assessment of air pollutant emissions from
  brick kilns,'' \emph{Atmospheric Environment}, vol.~98, pp. 549 -- 553, 2014.
  [Online]. Available:
  \url{http://www.sciencedirect.com/science/article/pii/S1352231014006888}
\BIBentrySTDinterwordspacing

\bibitem{rauf2022prospects}
A.~Rauf, S.~Shakir, A.~Ncube, H.~M. Abd-ur Rehman, A.~K. Janjua, S.~Khanum, and
  A.~H. Khoja, ``Prospects towards sustainability: A comparative study to
  evaluate the environmental performance of brick making kilns in pakistan,''
  \emph{Environmental Impact Assessment Review}, vol.~94, p. 106746, 2022.

\bibitem{he_optimal}
\BIBentryALTinterwordspacing
X.~He, X.~Xu, and Z.~Zheng, ``Optimal band analysis of a space-based
  multispectral sensor for urban air pollutant detection,'' \emph{Atmosphere},
  vol.~10, no.~10, p. 631, Oct 2019. [Online]. Available:
  \url{http://dx.doi.org/10.3390/atmos10100631}
\BIBentrySTDinterwordspacing

\bibitem{nazir2020kiln}
U.~Nazir, U.~K. Mian, M.~U. Sohail, M.~Taj, and M.~Uppal, ``Kiln-net: A gated
  neural network for detection of brick kilns in south asia,'' \emph{IEEE
  Journal of Selected Topics in Applied Earth Observations and Remote Sensing},
  vol.~13, pp. 3251--3262, 2020.

\end{thebibliography}

\end{document}